\documentclass[aps, prl, superscriptaddress,showkeys, preprint]{revtex4-1}
\usepackage{titlesec}
\usepackage{graphicx}
\usepackage{dcolumn}
\usepackage{bm}
\usepackage{color}
\usepackage{tabularx}
\usepackage{tabularray}
\usepackage{multirow, makecell, booktabs}
\usepackage{threeparttable}
\usepackage{array}
\usepackage{amsmath}
\usepackage{amssymb}
\usepackage{mhchem}
\usepackage{comment}
\usepackage{ragged2e} 

\usepackage{pbox}
\usepackage{float}
\DeclareUnicodeCharacter{0394}{$\Delta$}
\usepackage[colorlinks=true, linkcolor=blue, citecolor=blue, urlcolor=blue]{hyperref}

\usepackage{stmaryrd}  

\begin{document}


\preprint{}

\title{\textbf{Optical poling of a quantum ferroelectric metal across the order-disorder phase transition}}

\author{Mohamed Kandil}
\affiliation{Department of Physics, Auburn University, Auburn, AL 36849, USA}

\author{Yasuhide Tomioka}
\affiliation{National Institute of Advanced Industrial Science and Technology (AIST), Tsukuba 305-8565, Japan}

\author{Yafei Ren}
\affiliation{Department of Physics, University of Delaware, Newark, DE 19716, USA}

\author{Ryan Comes}
\affiliation{Department of Materials Science and Engineering, University of Delaware, Newark, DE 19716, USA}

\author{Isao H. Inoue}
\affiliation{National Institute of Advanced Industrial Science and Technology (AIST), Tsukuba 305-8565, Japan}

\author{Wencan Jin}
\email{wjin@auburn.edu}
\affiliation{Department of Physics, Auburn University, Auburn, AL 36849, USA}

\date{\today}

\begin{abstract}

\noindent Electron-doped strontium titanate has emerged as a prototypical quantum ferroelectric metal. It provides a fertile ground to explore how ferroelectric instability intertwined with itinerant electrons creates quantum phenomena, including unconventional superconductivity. Despite extensive studies, the microscopic origin of the ferroelectric transition remains unsettled, with distinct interpretations based on displacive mechanism driven by soft mode and order–disorder alignment of local dipoles. In particular, the local dipoles form nanoscale, spatially heterogeneous clusters, termed polar nanoregions, posing a significant challenge for probing or manipulating them. Here, using rotational anisotropy second harmonic generation, a symmetry-resolved probe, we quantify the orientational statistics of polar nanoregions in dilute electron-doped \ce{Sr_{0.95}Ba_{0.05}Ti_{1-x}Nb_{x}O3}. By tracking the alignment and meltdown of polar nanoregions in thermal cycles, we unambiguously demonstrate the order-disorder nature of the ferroelectric transition. We further show that, above transition temperature, femtosecond optical fields enable deterministic control of otherwise disordered polar nanoregions, realizing reversible write and readout of polar textures on ultrafast timescales. Our findings provide new insight into ferroelectric instability and establish an all-optical route of controlling polar metal systems where conventional electrical approaches are not feasible. 


\end{abstract}

\maketitle

The coexistence of polarity/ferroelectricity and metallicity in the same phase, as first discussed by Anderson and Blount in 1965 ~\cite{anderson1965symmetry}, has long been considered incompatible due to itinerant carrier screening of long-range dipolar interactions. Over the past decade, experimental realization of polar metal \ce{LiOsO3} \cite{shi2013ferroelectric} and switchable ferroelectric metal \ce{WTe2} \cite{fei2018ferroelectric}, has fueled the rapidly growing research interest in the design of a host of polar metals, which unlock the opportunites of exploring exotic quantum phenomena such as topology, multiferroics, complex magnetism, and unconventional superconductivity ~\cite{zhou2020review,bhowal2023polar,hickox2023polar}.

Quantum ferroelectric metals represent a unique and highly tunable polar metal system, in which metallicity develops in proximity to a ferroelectric quantum critical point (QCP) ~\cite{rowley2014ferroelectric,klein2023theory}. In contrast to canonical polar metals, which are characterized primarily by the polar structure coexisting with metallic conductivity, quantum ferroelectric metals arise from a quantum paraelectric parent compound. A prototypical example is electron-doped perovskite strontium titanate \ce{SrTiO3} (STO). In its stoichiometric form, STO is quantum paraelectric, whose ferroelectric phase transition is suppressed by quantum fluctuations~\cite{Müller1979, zhong1996effect, Gu2023quantum}. However, ferroelectricity can be induced by various non-thermal tuning parameters, such as isovalent doping ~\cite{bednorz1984sr, bianchi1995cluster, Lemanov1996, shirokov2006concentration}, oxygen isotope substitution ~\cite{itoh1999ferroelectricity, Blinc2008}, epitaxial strain~\cite{ahadi2019enhancing, Russell2019, Salmani2020, Salmani-Rezaie2020}, or optical excitation of lattice vibrations ~\cite{Nova2019, Li2019, fechner2024quenched}. Moreover, STO becomes metallic after dilute carrier doping by substitution of Sr with La, Ti with Nb, or introduction of oxygen vacancies ~\cite{Tomioka2022, Rischau2017}. In this regime, the emergence of superconductivity at exceptionally low carrier concentration reveals a non-Bardeen-Cooper-Schrieffer pairing mechanism ~\cite{lin2013fermi,ruhman2016superconductivity}. Remarkably, it has been shown that the superconducting transition temperature can be enhanced near the ferroelectric QCP~\cite{ahadi2019enhancing}, highlighting an intimate connection between the ferroelectric fluctuations, itinerant carriers, and unconventional superconductivity~\cite{collignon2019metallicity, gastiasoro2020superconductivity}.

Despite the vital importance of understanding the role of ferroelectricity in mediating superconductivity, the microscopic origin of ferroelectric transition remains under debate. Ferroelectric order in electron-doped STO is generally acknowledged as a displacive model ~\cite{Pertsev2000Phase,Antons2005Tunability,li2025classical}; as such, ferroelectricity can be predominantly induced by coherent photoexcitation of the soft mode using intense terahertz (THz) pulses and detected by optical second harmonic generation or x-ray diffuse scattering ~\cite{Nova2019,Li2019,fechner2024quenched}. Recently, growing experimental evidence reveals the order-disorder or relaxor characteristics of the ferroelectric transition in electron-doped STO and epitaxially strained \ce{KTaO3}, where long-range ferroelectric order arises from the alignment of preformed polar nanoregions ~\cite{Salmani2020,Salmani-Rezaie2020,Hallett2022,Zhu2024, schwaigert2026above}. In contrast to the displacive ferroelectrics, the local dipoles associated with the polar nanoregions tunnel between the two minimum-energy states of the double-well potential due to thermal or quantum fluctuations. However, direct access and manipulation of polar nanoregions are technically challenging. The metallic nature of electron-doped STO has largely precluded conventional electrical approaches. Moreover, fluctuating polar nanoregions produce vanishing macroscopic polarization on average ($\langle P\rangle=0$), rendering most experimental probes sensitive primarily to the existence of local polarity rather than their orientational statistics.

\begin{figure*}[ht!]
    \includegraphics[width=0.98\textwidth]{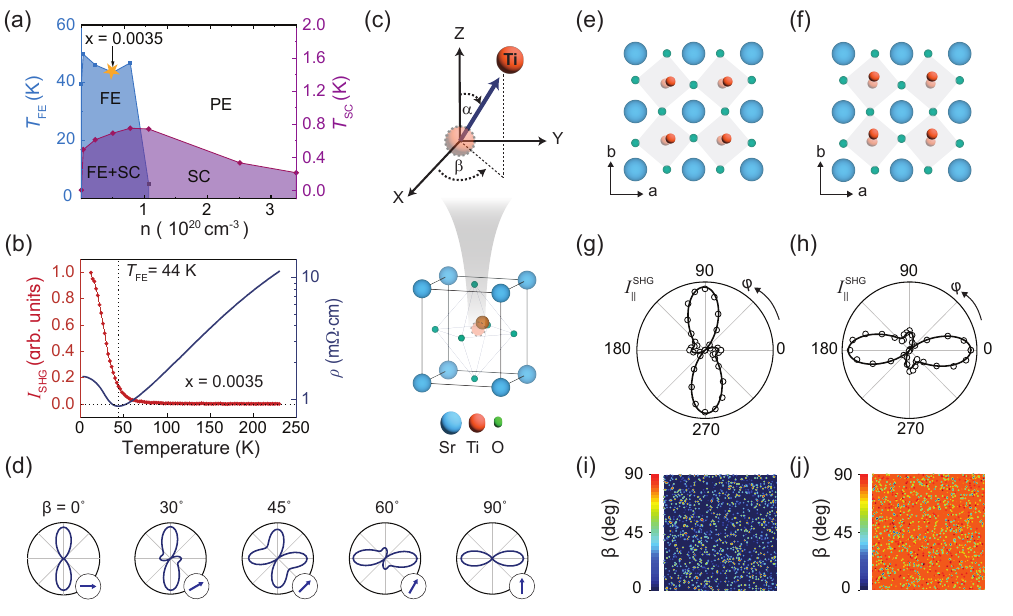}
    \caption{\label{fig:1} \textbf{Local dipoles induced SHG in the quantum ferroelectric metal.} (a) Temperature-carrier concentration phase diagram of the \ce{Sr_{0.95}Ba_{0.05}Ti_{1-x}Nb_{x}O3} bulk crystals. Abbreviations: PE, paraelectric; FE, ferroelectric; SC, superconducting. (b) Normalized SHG intensity (left axis) and electrical resistivity (right axis) as a function of temperature for doping concentration $x$ = 0.0035. (c) Displacement of the Ti atom off the center of the unit cell as illustrated by the polar vector in a spherical coordinate system. (d) Simulated RA-SHG patterns for $\beta$ = 0$^{\circ}$, 30$^{\circ}$, 45$^{\circ}$, 60$^{\circ}$, 90$^{\circ}$, respectively. (e,f) Schematic of polar nanoregions with minimum energy in the double-well potential. (g,h) Two types of RA-SHG patterns acquired through the spatial survey at 10 K. The patterns are fitted to the simulations of $100\times100$ cells with (i) $\beta$ is 10$^\circ$ in 80\% cells and randomly oriented in 20\% cells, and (j) $\beta$ is 80$^\circ$ in 80\% cells and randomly oriented in 20\% cells, respectively.}
\end{figure*}

Here, we present deterministic control of polar nanoregions in quantum ferroelectric metal \ce{Sr_{0.95}Ba_{0.05}Ti_{1-x}Nb_{x}O3} ($x$ = 0.0035) ~\cite{Tomioka2022}. The single-crystal samples provide a material platform free of epitaxial strain-induced ferroelectricity, and the synthesis method is described in Section I of the Supplemental Material ~\cite{SM}. The temperature-carrier concentration phase diagram (Fig.~\ref{fig:1}(a)) is constructed using resistivity and Hall mobility characterization ~\cite{Tomioka2022}, which presents the destruction of ferroelectricity and the emergence of a superconducting dome. To investigate the microscopic nature and optical manipulation of polar nanoregions, we choose Nb doping of $x$ = 0.0035, where ferroelectric order ($T_\mathrm{FE}$ = 44 K) coexists with superconductivity near the optimal critical temperature ($T_\mathrm{SC}$ = 0.69 K). Although 5\% Ba substitution of Sr suppresses antiferrodistortive (AFD) order featured by antiphase rotations of the oxygen octahedra, below the critical Ba concentration ($\sim$9.4\%), the AFD phase persists, within which local polarization arises from Ti displacements in the absence of long-range ferroelectricity ~\cite{menoret2002structural, Benedek2016,Bersuker2020}. Rotational anisotropy second harmonic generation (RA-SHG) ~\cite{Denev2011,ZHAO2018207, jin2020observation} probes symmetry-breaking phases of solids through nonlinear optical susceptibility tensor (see experimental method in Section I of Supplemental Material ~\cite{SM}). Beyond long-range orders, in the disordered regime, RA-SHG couples to $\langle P^2\rangle$-type local polarization fluctuations, enabling access to the orientational statistics of polar nanoregions. By tracking the evolution of RA-SHG patterns in consecutive thermal cycles, we quantitatively show the condensation of polar nanoregions into ferroelectric domains upon cooling and their melting upon warming, which unambiguously reveals the order-disorder nature of the ferroelectric transition. Strikingly, in the disordered phase, the orientations of the polar nanoregions can be aligned by femtosecond pump pulses whose frequency is off-resonant with any soft modes. We attribute this effect to an optical poling process, in which an optical field biasing the double-well potential allows to transiently imprint net polar orders along arbitrary orientations. Our work demonstrates an all-optical write and readout of local dipoles in a quantum ferroelectric metal. It paves the route of tuning ferroelectric fluctuations and unconventional superconductivity through light and promises applications such as ultrafast, optically addressable memory and programmable superconducting circuits. 
 
We first examine the origin of SHG response in a wide temperature range. As shown in Fig. ~\ref{fig:1}(b), the onset of the ferroelectric transition is confirmed by the resistivity anomaly at $T_\mathrm{FE}$ = 44 K ~\cite{Tomioka2022} (see additional data acquired at other carrier concentrations in Section II of Supplemental Material ~\cite{SM}). Below $T_\mathrm{FE}$, the SHG intensity rapidly increases, and above $T_\mathrm{FE}$, there exist residual SHG signals up to 230 K, which is a signature of preformed polar nanoregions in order-disorder transition. In previous report of thin film samples with compressive strain, the SHG signals across the ferroelectric phase transition were attributed to symmetry lowering from tetragonal 4/mmm (D$_\mathrm{4h}$) point group to the 4mm (C$_\mathrm{4v}$) subgroup ~\cite{Russell2019}. Yet in our case, the SHG signals are acquired in normal incidence geometry, in which electric dipole (ED) or electric quadrupole (EQ) SHG are absent in both point groups (See Section III of the Supplemental Material ~\cite{SM}). 

To account for the finite SHG signal above and below $T_\mathrm{FE}$, we develop a model based on the presence of local structural heterogeneity in a centrosymmetric lattice frame. As shown in Fig.~\ref{fig:1}(c), the local dipole associated with the off-center displacement of \ce{Ti} atom is characterized by a polar vector that features the polar angle off the $z$ axis ($\alpha$) and the azimuthal angle ($\beta$). The magnitude of the polar vector, i.e., the displacement length scale has been estimated at $\sim$ 10 pm in the strained STO thin films ~\cite{Salmani2020}. Therefore, we consider the polar vector as a perturbation to the tetragonal lattice structure. In the centrosymmetric tetragonal 4/mmm group, the ED contribution from the local dipoles can be expressed as 

\begin{equation}
\label{eq:1}
\mathbf{\chi}_{ijk}^{ED}(\mathbf{P}) =\sum\limits_{l} \frac{\partial \chi_{ijk}}{\partial P_l} P_{l} \text{ ,}
\end{equation}

\noindent where $\chi_{ijkl}^{EQ} =\frac{\partial\chi_{ijk}}{\partial P_l}$ is the electric quadrupole susceptibility tensor of the 4/mmm point group and $\mathbf{P} = ( \sin\alpha \cos\beta,\sin\alpha \sin\beta, \cos\alpha)$ is the local polar vector as defined in Fig.~\ref{fig:1}(c). In the parallel polarization configuration of normal incidence geometry, the second harmonic electric field from a single polar nanoregion is

\begin{equation}
\label{eq:2}
E_{||}^{2\omega}(\varphi) =  \sin(\alpha) 
[ A \, \sin(\beta - 3\varphi) + B \, \sin(\beta + \varphi)],
\end{equation}

\noindent where $\varphi$ is the azimuthal angle between the polarization of incident light and the $a$ axis of the sample; $A = (\chi_{xxxx} - \chi_{yxxy} - \chi_{yyxx} - \chi_{yxyx}$) and $B = (3\chi_{xxxx} + \chi_{yxxy} + \chi_{yyxx} + \chi_{yxyx}$) are linear combinations of $\chi_{ijkl}^{EQ}$ and remain constant under perturbation (see detailed procedure in Section IV of Supplemental Material ~\cite{SM}). The amplitude of the second harmonic electric field is proportional to $\sin{\alpha}$, while the angular anisotropy is determined by $\beta$. Figure~\ref {fig:1}(d) shows the simulated RA-SHG patterns for selected $\beta$ values in the 0$^\circ$-90$^\circ$ quadrant. These patterns exhibit only 2-fold rotational symmetry, indicating the presence of polar vector breaks vertical and diagonal mirror symmetries. Meanwhile, the patterns of $\beta=0^{\circ}$ and $\beta=90^{\circ}$ are still connected by C$_4$ operation, preserving the 4-fold rotational symmetry of the tetragonal lattice. 

At 10 K, we survey the sample at various locations using a laser beam size of 50 $\mu$m in diameter, which covers a substantial amount of polar nanoregions. We find that the SHG intensity shows little spatial variation, indicating the amplitude of Ti displacement (characterized by the in-plane projection sin$\alpha$) remains largely homogeneous. Meanwhile, our survey identifies two types of RA-SHG patterns with distinct angular anisotropy as shown in Figs.~\ref{fig:1}(g) and ~\ref{fig:1}(h). We consider the SHG response arising from the $\langle P^2\rangle$-type local polarization fluctuations 

\begin{equation}
I_{||}^{2\omega}(\varphi) =  \sin^2(\alpha) \sum_{n=1}^{N}[ A \, \sin(\beta_{n} - 3\varphi) + B \, \sin(\beta_{n} + \varphi)]^2,
\end{equation}

\noindent where $\beta_n$ is the orientation of the $n$th polar nanoregion. The RA-SHG patterns can thus be fitted through a simulation. We use $100\times100$ cells representing polar nanoregions within our beam spot, and in each cell, the polar orientation ($\beta$) is allowed to change randomly in the 0$^\circ$-90$^\circ$ range. The simulated RA-SHG pattern shows the best agreement with the experimental data when $\beta$ in $\sim80\%$ of the cells align with $10^{\circ}$ for Fig.~\ref{fig:1}(i) and $80^{\circ}$ for Fig.~\ref{fig:1}(j), respectively. $\beta$ in the rest $\sim20\%$ cells remains random, which indicates that the polar vectors are not fully ordered down to 10 K as reflected in the unsaturated SHG intensity shown in Fig.~\ref{fig:1}(b). The alignment of $\beta$ along $10^{\circ}$ and $80^{\circ}$ (equivalent to $-10^{\circ}$) are consistent with the \textit{ab initio} calculations, where the AFD rotation of $11^{\circ}$ gives rise to a minimum in the potential energy landscape ~\cite{fechner2024quenched}. Based on the observation and simulation, we have demonstrated that at 10 K the polar nanoregions condense into two types of ferroelectric-like domain states (see schematics in Fig.~\ref{fig:1}(e) and ~\ref{fig:1}(f)), whose orientations are determined by the interplay between the polar distortion and the AFD distortion ~\cite{aschauer2014competition,gu2018cooperative}. 

\begin{figure*}[ht!]
\includegraphics[width=0.98\textwidth]{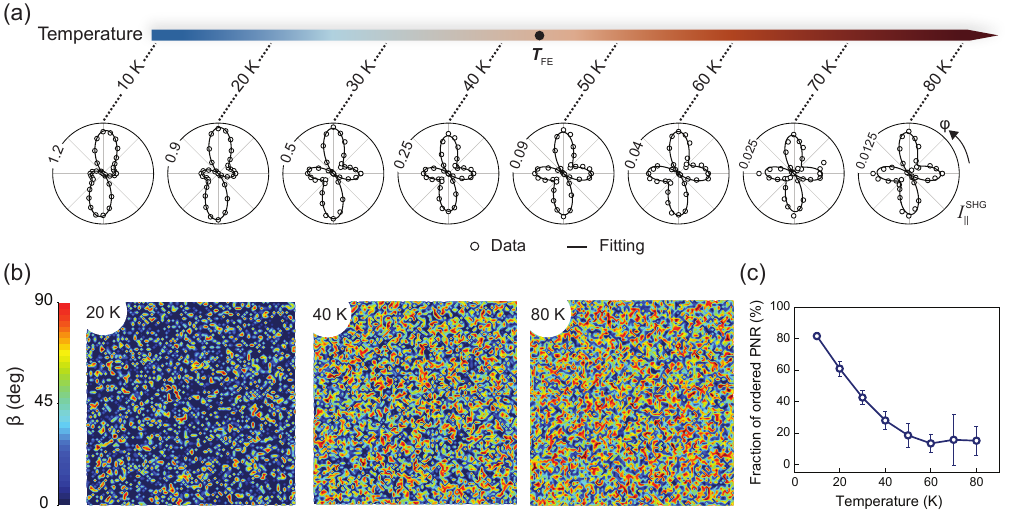}
\caption{\label{Fig:2} \textbf{Meltdown of ordered polar nanoregions in warmup cycle}. (a) Polar plots of RA-SHG data acquired in the parallel channel of the normal incidence configuration at selected temperatures below and above $T_\mathrm{FE}$. All data sets are normalized to a value of 1 corresponding to the maximum SHG intensity at 10 K. (b) Heatmaps ($100\times100$ cells) of simulated in-plane polarization orientation ($\beta$) that produce the best fit to the RA-SHG data of 20 K, 40 K, and 80 K, respectively. (c) Population of polar nanoregions (PNR) with $\beta$ value of $10^{\circ}$ as a function of temperature. Error bars represent 95\% confidence interval of fits to the RA-SHG patterns.}
\end{figure*}

We then warm up the sample while tracking the evolution of RA-SHG patterns acquired from the domain state having 80\% of $\beta=10^{\circ}$. As shown in Fig.~\ref{Fig:2}(a), from 10 K to 80 K, the SHG intensity decreases by two orders of magnitude.  Figure~\ref{Fig:2}(b) displays the heatmaps of $\beta$ values in $100\times100$ cells that produce the best fit to the RA-SHG patterns at 20 K, 40 K, and 80 K, respectively. The heatmaps show the growing randomness of $\beta$ with increasing temperature. Specifically, from 10 K to 80 K, the population of the polar nanoregions with $\beta$ aligned along $10^{\circ}$ decreases from 80\% to 10\% (see Fig.~\ref{Fig:2}(c)). Similar results are also obtained from the domain state where $\beta=-10^{\circ}$ (see Section IV of Supplemental Material ~\cite{SM}). Up to here, we have established the order-disorder nature of the ferroelectric transition. The randomly oriented polar nanoregions exist well above $T_\mathrm{FE}$. They align to the minimum-energy states at the lowest temperature, forming ferroelectric-like domain states, and melt down upon warming up.

\begin{figure*}[ht!]
\includegraphics[width=0.98\textwidth]{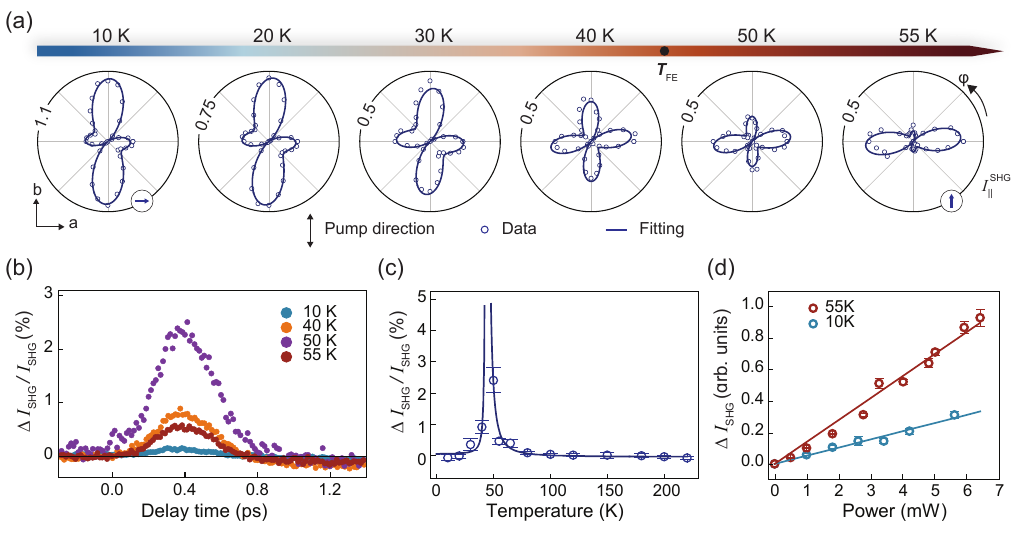}
\caption{\label{Fig:3} \textbf{Dynamics of polar nanoregions across the order-disorder transition}. (a) RA-SHG patterns acquired in a warm-up cycle under pump pulses with linear polarization along $b$ axis (see pump direction denoted as a double arrow). (b) Relative SHG change ($\Delta I_\mathrm{SHG}/I_\mathrm{SHG}$) as a function of delay time at selected temperatures. (c) Maximum $\Delta I_\mathrm{SHG}/I_\mathrm{SHG}$ as a function of temperature. Solid curves are fits to $|T-T_\mathrm{FE}|^{-\gamma}$. (d) Comparison between the relative change of SHG intensity as a function of pump power at $10\text{\,K}$ and $55\text{\,K}$. Error bars represent 95\% confidence interval of fits to the change of SHG intensity.}
\end{figure*}

We then investigate the dynamics of the polar nanoregions in a pump-probe experiment. At 10 K, we let the pump (700 nm) and probe (800 nm) beams spatially overlap at the domain states where 80\% $\beta$ aligns along 10$^{\circ}$ (Fig.~\ref{fig:1}(e)). Polarization of the pump pulse is set along the $b$ axis, and RA-SHG patterns are acquired using a probe pulse with 0.4 ps delay, at which the pump-induced change in SHG intensity becomes maximum (see Fig.~\ref{Fig:3}(b)). As shown in the RA-SHG patterns Fig.~\ref{Fig:3}(a), from 10 K to 55 K, the ordered polar nanoregions melt down, and then realign along the polarization direction of the pump pulse. We highlight several key features in this process. First, the pump laser frequency is far apart from any soft polar mode; thus, ferroelectricity induced by the coherent phonon excitation does not apply. Second, the response time of the pump-induced change in SHG intensity is $\sim$1 ps (see Fig.~\ref{Fig:3}(b)), which is consistent with the ferroelectric transition temperature of 44 K that corresponds to $\sim$4 meV depth of the double-well potential. Third, the pump-induced SHG intensity change ($\Delta I_\mathrm{SHG}$) diverges as the temperature approaches $T_\mathrm{FE}$, which can be fitted to the susceptibility behavior $|T-T_\mathrm{FE}|^{-\gamma}$. The critical exponent $\gamma$ of 0.99 $\pm$ 0.11 is consistent with a minimal model of the Curie-Weiss law (see Fig.~\ref{Fig:3}(c)). Lastly and most importantly, as shown in Fig.~\ref{Fig:3}(d), $\Delta I_\mathrm{SHG}$ grows linearly with the increasing pump power. This is distinct from the typical electric field-induced second harmonic, in which $\Delta I_\mathrm{SHG}$ is proportional to the pump field strength ~\cite{Cheng2023,xsrf-t1hj}. 

To address this discrepancy, we consider the polarization modulation induced by a pump pulse $E_\mathrm{pump}^{\omega}$. It is commonly assumed that the second-order susceptibility changes as $\chi^{(2)}\rightarrow\chi^{(2)}+\Delta\chi^{(2)}$, with $\Delta\chi^{(2)}\propto E_\mathrm{pump}^{\omega}$, and thus the resulting change in the SHG intensity is 

\begin{equation}
\label{eq:4}
\Delta I_\mathrm{SHG} \propto [\chi^{(2)}+\Delta\chi^{(2)}]^2-[\chi^{(2)}]^2 = 2\chi^{(2)}\Delta\chi^{(2)}+[\Delta\chi^{(2)}]^2
\end{equation}

\noindent For a system with long-range polar order, the $2\chi^{(2)}\Delta\chi^{(2)}$ term dominates such that $\Delta I_\mathrm{SHG} \approx 2\chi^{(2)}\Delta\chi^{(2)} \propto E_\mathrm{pump}^{\omega}$.
In contrast, for dilute electron-doped STO, when the polar nanoregions are randomly oriented in the disordered phase at 55 K, the global polarization characterized by $\chi^{(2)}$ is zero. Meanwhile, in the vicinity of $T_\mathrm{FE}$, the local dipoles are subject to strong fluctuations as evidenced by the diverging susceptibility (see Fig.~\ref{Fig:3}(c)). As a result, $\Delta I_\mathrm{SHG}$ is determined by the $[\Delta\chi^{(2)}]^2$ term, which is proportional to $[E_\mathrm{pump}^{\omega}]^2$, that is, the pump power. Our results resemble an optical poling process based on $\langle P^2\rangle$-type local polarization correlation, despite macroscopic polarization $\langle P\rangle$ remaining zero. When the local polarization fluctuation is suppressed in the ordered phase at 10 K, the optical poling becomes much less efficient as manifested in the substantially lower slope compared to 55 K (see Fig.~\ref{Fig:3}(d)). 

It is worth noting that for dilute electron-doped STO (carrier concentration $n = 5\times 10^{19}$ cm$^{-3}$, effective mass $m^*\sim5m_e$, high frequency dielectric constant $\epsilon_\infty=5.5$), the plasma frequency estimated using the Drude model is $\sim$12 THz. Since the fundamental and second harmonic laser frequencies are much higher than the plasma frequency, the laser beam can penetrate deep into the bulk of the crystal, allowing it to tune and probe internal polar nanodomains rather than just the surface.

\begin{figure*}[ht!]
\includegraphics[width=0.98\textwidth]{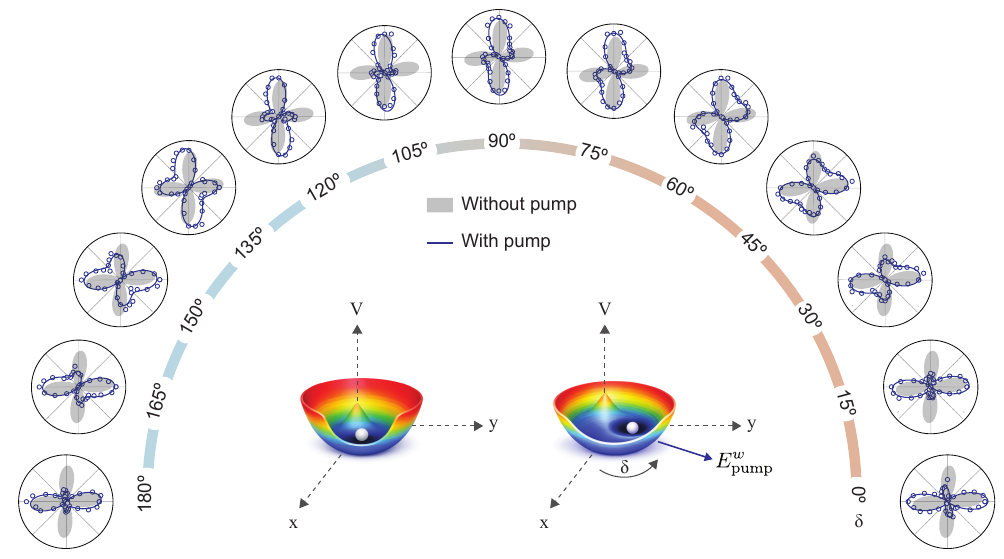}
\caption{\label{Fig:4} \textbf{Optical poling of polar nanoregions in the disordered phase}. RA-SHG patterns acquired with optical pump from varying polarization directions ($\delta$) at 55 K. Blue empty circles are the raw data, and blue curves are best fits to the simulation of $100\times100$ cells. The RA-SHG patterns without pump at 55 K (gray shade) are overlaid for comparison. The inset shows (left) the isotropic potential landscape when polar nanoregions are randomly oriented in the disordered phase, and (right) the poling field $E_\mathrm{pump}^{\omega}$ distorts the potential well and creates a new minimum.} 
\end{figure*}

Finally, we show that the poling field can control the preferred orientation of the polar nanoregions in the disordered phase. At 55 K, we tune the polarization direction ($\delta$) of the pump pulse from $0^\circ$ to $180^\circ$ in $15^\circ$ steps. For each $\delta$, we acquire an RA-SHG pattern using the probe beam. As shown in Fig.~\ref{Fig:4}, the orientation of polar nanoregions ($\beta$) aligns with the pump pulse polarization, indicating that the optical field applies a directional distortion to the potential well landscape (see inset of Fig.~\ref{Fig:4}). Also, our simulation of the pump-probe RA-SHG patterns shows that over 80\% of the polar nanoregions are consistently aligned with the optical field (see Section VI of the Supplemental Material ~\cite{SM}), which demonstrates satisfactory optical poling efficiency. Such an all-optical write and readout protocol for polar textures in a quantum ferroelectric metal opens up opportunities for the development of ultrafast memory devices and programmable superconducting circuits where the superconducting transition temperature can be tuned through light-induced ferroelectric fluctuations. We envision that systematic ultrafast optical studies as a function of carrier concentrations approaching the putative QCP may offer new insight into how critical fluctuations impact superconductivity and nonequilibrium dynamics. Furthermore, the ability to fully manipulate the orientation of local dipoles across the order-disorder transition paves the way for exploring exotic topological phenomena and unconventional pairing mechanisms in a wide range of polar metals.

\section{Acknowledgments}

\noindent We acknowledge technical assistance from Chunli Tang and Hussam Mustafa. Also, we thank Liuyan Zhao, Yinong Zhou, Xiaoyu Guo, and Xianghan Xu for helpful discussions. W. J. and R.C. acknowledge support by the Air Force Office of Scientific Research under Award No. FA9550-231-0499 and the DOE Office of Science Grant No. DE-SC0023478. Y.T. and I.H.I. were supported by JSPS KAKENHI Grant Numbers 19H01844 and 25H00451.

\bibliographystyle{naturemag}
\bibliography{references.bib}

@article{Tomioka2022,
   author = {Tomioka, Yasuhide and Shirakawa, Naoki and Inoue, Isao H.},
   title = {Superconductivity enhancement in polar metal regions of \ce{Sr_{0.95}Ba_{0.05}TiO3} and \ce{Sr_{0.985}Ca_{0.015}TiO3} revealed by systematic \ce{Nb} doping},
   journal = {npj Quantum Materials},
   volume = {7},
   number = {1},
   pages = {111},
   ISSN = {2397-4648},
   DOI = {10.1038/s41535-022-00524-9},
   url = {https://doi.org/10.1038/s41535-022-00524-9},
   year = {2022},
   type = {Journal Article}
}

@article{jin2020observation,
  title={Observation of a ferro-rotational order coupled with second-order nonlinear optical fields},
  author={Jin, Wencan and Drueke, Elizabeth and Li, Siwen and Admasu, Alemayehu and Owen, Rachel and Day, Matthew and Sun, Kai and Cheong, Sang-Wook and Zhao, Liuyan},
  journal={Nature Physics},
  volume={16},
  number={1},
  pages={42--46},
  url={https://doi.org/10.1038/s41567-019-0695-1},
  year={2020},
  publisher={Nature Publishing Group UK London}
}

@article{anderson1965symmetry,
  title = {Symmetry considerations on martensitic transformations:`` ferroelectric" metals?},
  author = {Anderson, P. W. and Blount, E. I.},
  journal = {Physical Review Letters},
  volume = {14},
  issue = {7},
  pages = {217--219},
  numpages = {0},
  year = {1965},
  publisher = {American Physical Society},
  doi = {10.1103/PhysRevLett.14.217},
  url = {https://link.aps.org/doi/10.1103/PhysRevLett.14.217}
}

@article{shi2013ferroelectric,
  title={A ferroelectric-like structural transition in a metal},
  author={Shi, Youguo and Guo, Yanfeng and Wang, Xia and Princep, Andrew J and Khalyavin, Dmitry and Manuel, Pascal and Michiue, Yuichi and Sato, Akira and Tsuda, Kenji and Yu, Shan and others},
  journal={Nature Materials},
  volume={12},
  number={11},
  pages={1024--1027},
  url ={https://doi.org/10.1038/nmat3754},
  year={2013},
  publisher={Nature Publishing Group UK London}
}

@article{fei2018ferroelectric,
  title={Ferroelectric switching of a two-dimensional metal},
  author={Fei, Zaiyao and Zhao, Wenjin and Palomaki, Tauno A and Sun, Bosong and Miller, Moira K and Zhao, Zhiying and Yan, Jiaqiang and Xu, Xiaodong and Cobden, David H},
  journal={Nature},
  volume={560},
  number={7718},
  pages={336--339},
  url={https://doi.org/10.1038/s41586-018-0336-3},
  year={2018},
  publisher={Nature Publishing Group UK London}
}

@article{zhou2020review,
  title={Review on ferroelectric/polar metals},
  author={Zhou, WX and Ariando, Ariando},
  journal={Japanese Journal of Applied Physics},
  volume={59},
  number={SI},
  pages={SI0802},
  url={https://doi.org/10.35848/1347-4065/ab8bbf},
  year={2020},
  publisher={IOP Publishing}
}

@article{bhowal2023polar,
  title={Polar metals: principles and prospects},
  author={Bhowal, Sayantika and Spaldin, Nicola A},
  journal={Annual Review of Materials Research},
  volume={53},
  number={1},
  pages={53--79},
  url={https://doi.org/10.1146/annurev-matsci-080921-105501},
  year={2023},
  publisher={Annual Reviews}
}

@article{hickox2023polar,
  title={Polar metals taxonomy for materials classification and discovery},
  author={Hickox-Young, Daniel and Puggioni, Danilo and Rondinelli, James M},
  journal={Physical Review Materials},
  volume={7},
  number={1},
  pages={010301},
  url={https://doi.org/10.1103/PhysRevMaterials.7.010301},
  year={2023},
  publisher={APS}
}

@article{Müller1979,
   author = {Müller, K. A. and Burkard, H.},
   title = {\ce{SrTiO3}: An intrinsic quantum paraelectric below 4 {K}},
   journal = {Physical Review B},
   volume = {19},
   number = {7},
   pages = {3593-3602},
   DOI = {10.1103/PhysRevB.19.3593},
   url = {https://link.aps.org/doi/10.1103/PhysRevB.19.3593},
   year = {1979},
   type = {Journal Article}
}

@article{zhong1996effect,
  title={Effect of quantum fluctuations on structural phase transitions in \ce{SrTiO3} and \ce{BaTiO3}},
  author={Zhong, W and Vanderbilt, David},
  journal={Physical Review B},
  volume={53},
  number={9},
  pages={5047},
  year={1996},
  publisher={APS},
  url={https://doi.org/10.1103/PhysRevB.53.5047}
}

@article{Gu2023quantum,
   author = {Gu, Fangyuan and Wang, Jie and Lang, Zi-Jian and Ku, Wei},
   title = {Quantum fluctuation of ferroelectric order in polar metals},
   journal = {npj Quantum Materials},
   volume = {8},
   number = {1},
   pages = {49},
   ISSN = {2397-4648},
   DOI = {10.1038/s41535-023-00578-3},
   url = {https://doi.org/10.1038/s41535-023-00578-3},
   year = {2023},
   type = {Journal Article}
}

@article{bednorz1984sr,
  title={\ce{Sr_{1-x}Ca_{x}TiO3}: an {XY} quantum ferroelectric with transition to randomness},
  author={Bednorz, JG and M{\"u}ller, KA},
  journal={Physical Review Letters},
  volume={52},
  number={25},
  pages={2289},
  year={1984},
  publisher={APS},
  url={https://doi.org/10.1103/PhysRevLett.52.2289}
}

@article{bianchi1995cluster,
  title={Cluster and domain-state dynamics of ferroelectric \ce{Sr_{1-x}Ca_{x}TiO3} (x= 0.007)},
  author={Bianchi, U and Dec, J and Kleemann, W and Bednorz, JG},
  journal={Physical Review B},
  volume={51},
  number={14},
  pages={8737},
  year={1995},
  publisher={APS},
  url={https://doi.org/10.1103/PhysRevB.51.8737}
}

@article{Lemanov1996,
   author = {Lemanov, V. V. and Smirnova, E. P. and Syrnikov, P. P. and Tarakanov, E. A.},
   title = {Phase transitions and glasslike behavior in \ce{Sr_{1-x}Ba_{x}TiO3}},
   journal = {Physical Review B},
   volume = {54},
   number = {5},
   pages = {3151-3157},
   DOI = {10.1103/PhysRevB.54.3151},
   url = {https://link.aps.org/doi/10.1103/PhysRevB.54.3151},
   year = {1996},
   type = {Journal Article}
}

@article{shirokov2006concentration,
  title={Concentration phase diagram of \ce{Ba_{x}Sr_{1-x}TiO3} solid solutions},
  author={Shirokov, VB and Torgashev, VI and Bakirov, AA and Lemanov, VV},
  journal={Physical Review B},
  volume={73},
  number={10},
  pages={104116},
  year={2006},
  publisher={APS},
  url={https://doi.org/10.1103/PhysRevB.73.104116}
}

@article{gastiasoro2020superconductivity,
  title={Superconductivity in dilute \ce{SrTiO3}: A review},
  author={Gastiasoro, Maria N and Ruhman, Jonathan and Fernandes, Rafael M},
  journal={Annals of Physics},
  volume={417},
  pages={168107},
  url={https://doi.org/10.1016/j.aop.2020.168107},
  year={2020},
  publisher={Elsevier}
}

@article{collignon2019metallicity,
  title={Metallicity and superconductivity in doped strontium titanate},
  author={Collignon, Cl{\'e}ment and Lin, Xiao and Rischau, Carl Willem and Fauqu{\'e}, Beno{\^\i}t and Behnia, Kamran},
  journal={Annual Review of Condensed Matter Physics},
  volume={10},
  number={1},
  pages={25--44},
  url={https://doi.org/10.1146/annurev-conmatphys-031218-013144},
  year={2019},
  publisher={Annual Reviews}
}

@article{ahadi2019enhancing,
  title={Enhancing superconductivity in \ce{SrTiO3} films with strain},
  author={Ahadi, Kaveh and Galletti, Luca and Li, Yuntian and Salmani-Rezaie, Salva and Wu, Wangzhou and Stemmer, Susanne},
  journal={Science advances},
  volume={5},
  number={4},
  pages={eaaw0120},
  url={https://doi.org/10.1126/sciadv.aaw0120},
  year={2019},
  publisher={American Association for the Advancement of Science}
}

@article{fechner2024quenched,
  title={Quenched lattice fluctuations in optically driven \ce{SrTiO3}},
  author={Fechner, M and F{\"o}rst, M and Orenstein, G and Krapivin, V and Disa, AS and Buzzi, M and Von Hoegen, A and De La Pena, G and Nguyen, QL and Mankowsky, R and others},
  journal={Nature Materials},
  volume={23},
  number={3},
  pages={363--368},
  year={2024},
  publisher={Nature Publishing Group UK London},
  url={https://doi.org/10.1038/s41563-023-01791-y}
}

@article{aschauer2014competition,
  title={Competition and cooperation between antiferrodistortive and ferroelectric instabilities in the model perovskite \ce{SrTiO3}},
  author={Aschauer, Ulrich and Spaldin, Nicola A},
  journal={Journal of Physics: Condensed Matter},
  volume={26},
  number={12},
  pages={122203},
  year={2014},
  publisher={IOP Publishing},
  url={https://iopscience.iop.org/article/10.1088/0953-8984/26/12/122203}
}

@article{gu2018cooperative,
  title={Cooperative couplings between octahedral rotations and ferroelectricity in perovskites and related materials},
  author={Gu, Teng and Scarbrough, Timothy and Yang, Yurong and {\'I}{\~n}iguez, Jorge and Bellaiche, L and Xiang, HJ},
  journal={Physical Review Letters},
  volume={120},
  number={19},
  pages={197602},
  year={2018},
  publisher={APS},
  url={https://doi.org/10.1103/PhysRevLett.120.197602}
}

@article{menoret2002structural,
  title={Structural evolution and polar order in \ce{Sr_{1-x}Ba_xTiO3}},
  author={Menoret, C and Kiat, Jean-Michel and Dkhil, Brahim and Dunlop, M and Dammak, Hichem and Hernandez, O},
  journal={Physical Review B},
  volume={65},
  number={22},
  pages={224104},
  year={2002},
  publisher={APS},
  url={https://doi.org/10.1103/PhysRevB.65.224104}
}

@article{Blinc2008,
   author = {Blinc, R. and Laguta, V. V. and Zalar, B. and Itoh, M. and Krakauer, H.},
   title = {\ce{$^{17}$O} quadrupole coupling and the origin of ferroelectricity in isotopically enriched \ce{BaTiO3} and \ce{SrTiO3}},
   journal = {Journal of Physics: Condensed Matter},
   volume = {20},
   number = {8},
   pages = {085204},
   ISSN = {0953-8984},
   DOI = {10.1088/0953-8984/20/8/085204},
   url = {https://dx.doi.org/10.1088/0953-8984/20/8/085204},
   year = {2008},
   type = {Journal Article}
}

@article{Benedek2016,
   author = {Benedek, Nicole A. and Birol, Turan},
   title = {‘{Ferroelectric}’ metals reexamined: fundamental mechanisms and design considerations for new materials},
   journal = {Journal of Materials Chemistry C},
   volume = {4},
   number = {18},
   pages = {4000-4015},
   ISSN = {2050-7526},
   DOI = {10.1039/C5TC03856A},
   url = {http://dx.doi.org/10.1039/C5TC03856A},
   year = {2016},
   type = {Journal Article}
}

@article{Rischau2017,
   author = {Rischau, Carl Willem and Lin, Xiao and Grams, Christoph P. and Finck, Dennis and Harms, Steffen and Engelmayer, Johannes and Lorenz, Thomas and Gallais, Yann and Fauqué, Benoît and Hemberger, Joachim and Behnia, Kamran},
   title = {A ferroelectric quantum phase transition inside the superconducting dome of \ce{Sr_{1-x}Ca_{x}TiO_{3-\delta}}},
   journal = {Nature Physics},
   volume = {13},
   number = {7},
   pages = {643-648},
   ISSN = {1745-2481},
   DOI = {10.1038/nphys4085},
   url = {https://doi.org/10.1038/nphys4085},
   year = {2017},
   type = {Journal Article}
}

@article{rowley2014ferroelectric,
  title={Ferroelectric quantum criticality},
  author={Rowley, SE and Spalek, LJ and Smith, RP and Dean, MPM and Itoh, M and Scott, JF and Lonzarich, GG and Saxena, SS},
  journal={Nature Physics},
  volume={10},
  number={5},
  pages={367--372},
  url={https://doi.org/10.1038/nphys2924},
  year={2014},
  publisher={Nature Publishing Group UK London}
}

@article{Russell2019,
   author = {Russell, Ryan and Ratcliff, Noah and Ahadi, Kaveh and Dong, Lianyang and Stemmer, Susanne and Harter, John W.},
   title = {Ferroelectric enhancement of superconductivity in compressively strained \ce{SrTiO3} films},
   journal = {Physical Review Materials},
   volume = {3},
   number = {9},
   pages = {091401},
   DOI = {10.1103/PhysRevMaterials.3.091401},
   url = {https://link.aps.org/doi/10.1103/PhysRevMaterials.3.091401},
   year = {2019},
   type = {Journal Article}
}

@article{Li2019,
   author = {Li, Xian and Qiu, Tian and Zhang, Jiahao and Baldini, Edoardo and Lu, Jian and Rappe, Andrew M. and Nelson, Keith A.},
   title = {Terahertz field induced ferroelectricity in quantum paraelectric \ce{SrTiO3}},
   journal = {Science},
   volume = {364},
   number = {6445},
   pages = {1079-1082},
   DOI = {doi:10.1126/science.aaw4913},
   url = {https://www.science.org/doi/abs/10.1126/science.aaw4913},
   year = {2019},
   type = {Journal Article}
}

@article{itoh1999ferroelectricity,
  title={Ferroelectricity induced by oxygen isotope exchange in strontium titanate perovskite},
  author={Itoh, M and Wang, R and Inaguma, Y and Yamaguchi, T and Shan, YJ and Nakamura, T},
  journal={Physical Review Letters},
  volume={82},
  number={17},
  pages={3540},
  url={https://doi.org/10.1103/PhysRevLett.82.3540},
  year={1999},
  publisher={APS}
}

@article{Nova2019,
   author = {Nova, T. F. and Disa, A. S. and Fechner, M. and Cavalleri, A.},
   title = {Metastable ferroelectricity in optically strained \ce{SrTiO3}},
   journal = {Science},
   volume = {364},
   number = {6445},
   pages = {1075-1079},
   DOI = {doi:10.1126/science.aaw4911},
   url = {https://www.science.org/doi/abs/10.1126/science.aaw4911},
   year = {2019},
   type = {Journal Article}
}

@article{Salmani-Rezaie2020,
   author = {Salmani-Rezaie, Salva and Ahadi, Kaveh and Stemmer, Susanne},
   title = {Polar Nanodomains in a Ferroelectric Superconductor},
   journal = {Nano Letters},
   volume = {20},
   number = {9},
   pages = {6542-6547},
   ISSN = {1530-6984},
   DOI = {10.1021/acs.nanolett.0c02285},
   url = {https://doi.org/10.1021/acs.nanolett.0c02285},
   year = {2020},
   type = {Journal Article}
}

@article{Salmani2020,
   author = {Salmani-Rezaie, Salva and Ahadi, Kaveh and Strickland, William M and Stemmer, Susanne},
   title = {Order-Disorder Ferroelectric Transition of Strained \ce{SrTiO3}},
   journal = {Physical Review Letters},
   volume = {125},
   number = {8},
   pages = {087601},
   DOI = {10.1103/PhysRevLett.125.087601},
   url = {https://link.aps.org/doi/10.1103/PhysRevLett.125.087601},
   year = {2020},
   type = {Journal Article}
}

@article{Zhu2024,
   author = {Zhu, Guomin and Hallett, Alex and Combs, Nicholas G. and Jeong, Hanbyeol and Genc, Arda and Harter, John W. and Stemmer, Susanne},
   title = {Coexistence of antiferrodistortive and polar order in a superconducting \ce{SrTiO3} film},
   journal = {Physical Review Materials},
   volume = {8},
   number = {5},
   pages = {L051801},
   DOI = {10.1103/PhysRevMaterials.8.L051801},
   url = {https://link.aps.org/doi/10.1103/PhysRevMaterials.8.L051801},
   year = {2024},
   type = {Journal Article}
}

@article{lin2013fermi,
  title={Fermi surface of the most dilute superconductor},
  author={Lin, Xiao and Zhu, Zengwei and Fauqu{\'e}, Beno{\^\i}t and Behnia, Kamran},
  journal={Physical Review X},
  volume={3},
  number={2},
  pages={021002},
  year={2013},
  publisher={APS},
  url={https://doi.org/10.1103/PhysRevX.3.021002}
}

@article{Bersuker2020,
   author = {Bersuker, Isaac B. and Polinger, Victor},
   title = {Perovskite Crystals: Unique Pseudo-Jahn–Teller Origin of Ferroelectricity, Multiferroicity, Permittivity, Flexoelectricity, and Polar Nanoregions},
   journal = {Condensed Matter},
   volume = {5},
   number = {4},
   pages = {68},
   ISSN = {2410-3896},
   url = {https://www.mdpi.com/2410-3896/5/4/68},
   year = {2020},
   type = {Journal Article}
}

@article{klein2023theory,
  title={Theory of criticality for quantum ferroelectric metals},
  author={Klein, Avraham and Kozii, Vladyslav and Ruhman, Jonathan and Fernandes, Rafael M},
  journal={Physical Review B},
  volume={107},
  number={16},
  pages={165110},
  year={2023},
  publisher={APS},
  url={https://doi.org/10.1103/PhysRevB.107.165110}
}

@article{ruhman2016superconductivity,
  title={Superconductivity at very low density: The case of strontium titanate},
  author={Ruhman, Jonathan and Lee, Patrick A},
  journal={Physical Review B},
  volume={94},
  number={22},
  pages={224515},
  year={2016},
  publisher={APS},
  url={https://doi.org/10.1103/PhysRevB.94.224515}
}

@article{Hallett2022,
   author = {Hallett, Alex and Harter, John W.},
   title = {Modeling polar order in compressively strained \ce{SrTiO3}},
   journal = {Physical Review B},
   volume = {106},
   number = {21},
   pages = {214107},
   DOI = {10.1103/PhysRevB.106.214107},
   url = {https://link.aps.org/doi/10.1103/PhysRevB.106.214107},
   year = {2022},
   type = {Journal Article}
}

@article{Cheng2023,
   author = {Cheng, Bing and Kramer, Patrick L. and Shen, Zhi-Xun and Hoffmann, Matthias C.},
   title = {Terahertz-Driven Local Dipolar Correlation in a Quantum Paraelectric},
   journal = {Physical Review Letters},
   volume = {130},
   number = {12},
   pages = {126902},
   DOI = {10.1103/PhysRevLett.130.126902},
   url = {https://link.aps.org/doi/10.1103/PhysRevLett.130.126902},
   year = {2023},
   type = {Journal Article}
}

@article{xsrf-t1hj,
  title = {Temporal modulation of second harmonic generation in ferroelectrics by a pulsed electric field},
  author = {Ono, Atsushi},
  journal = {Physical Review B},
  volume = {112},
  issue = {11},
  pages = {115146},
  url = {https://link.aps.org/doi/10.1103/xsrf-t1hj},
  year = {2025},
  publisher = {American Physical Society}
}

@article{schwaigert2026above,
  title={Above Room Temperature Ferroelectricity in Epitaxially Strained \ce{KTaO3}},
  author={Schwaigert, Tobias and Salmani-Rezaie, Salva and Hazra, Sankalpa and Saha, Utkarsh and Ramesh, Maya and Ross, Aiden and Pamuk, Betul and Chen, Long-Qing and Muller, David A and Schlom, Darrell G and others},
  journal={arXiv preprint arXiv:2601.14627},
  url ={https://doi.org/10.48550/arXiv.2601.14627},
  year={2026}
}

@article{Pertsev2000Phase,
  title = {Phase transitions and strain-induced ferroelectricity in \ce{SrTiO3} epitaxial thin films},
  author = {Pertsev, N. A. and Tagantsev, A. K. and Setter, N.},
  journal = {Physical Review B},
  volume = {61},
  issue = {2},
  pages = {R825--R829},
  numpages = {0},
  year = {2000},
  month = {Jan},
  publisher = {American Physical Society},
  doi = {10.1103/PhysRevB.61.R825},
  url = {https://link.aps.org/doi/10.1103/PhysRevB.61.R825}
}

@article{Antons2005Tunability,
  title = {Tunability of the dielectric response of epitaxially strained \ce{SrTiO3} from first principles},
  author = {Antons, Armin and Neaton, J. B. and Rabe, Karin M. and Vanderbilt, David},
  journal = {Physical Review B},
  volume = {71},
  issue = {2},
  pages = {024102},
  numpages = {11},
  year = {2005},
  month = {Jan},
  publisher = {American Physical Society},
  doi = {10.1103/PhysRevB.71.024102},
  url = {https://link.aps.org/doi/10.1103/PhysRevB.71.024102}
}

@article{li2025classical,
  title={The classical-to-quantum crossover in the strain-induced ferroelectric transition in \ce{SrTiO3} membranes},
  author={Li, Jiarui and Lee, Yonghun and Choi, Yongseong and Kim, Jong-Woo and Thompson, Paul and Crust, Kevin J and Xu, Ruijuan and Hwang, Harold Y and Ryan, Philip J and Lee, Wei-Sheng},
  journal={Nature Communications},
  volume={16},
  number={1},
  pages={4445},
  year={2025},
  publisher={Nature Publishing Group UK London},
  url={https://doi.org/10.1038/s41467-025-59517-4}
}

@article{Denev2011,
   author = {Denev, Sava A. and Lummen, Tom T. A. and Barnes, Eftihia and Kumar, Amit and Gopalan, Venkatraman},
   title = {Probing Ferroelectrics Using Optical Second Harmonic Generation},
   journal = {Journal of the American Ceramic Society},
   volume = {94},
   number = {9},
   pages = {2699-2727},
   ISSN = {0002-7820},
   url = {https://ceramics.onlinelibrary.wiley.com/doi/abs/10.1111/j.1551-2916.2011.04740.x},
   year = {2011},
   type = {Journal Article}
}

@article{ZHAO2018207,
title = {Second Harmonic Generation Spectroscopy of Hidden Phases},
editor = {Bob D. Guenther and Duncan G. Steel},
booktitle = {Encyclopedia of Modern Optics (Second Edition)},
publisher = {Elsevier},
edition = {Second Edition},
address = {Oxford},
pages = {207-226},
year = {2018},
isbn = {978-0-12-814982-9},
url={https://www.sciencedirect.com/science/article/pii/B9780128035818095333},
author = {Liuyan Zhao and Darius Torchinsky and John Harter and Alberto {de la Torre} and David Hsieh}
}

@misc{SM,
  note = "See Supplemental Material for synthesis and experimental methods, sample characterization, procedures of SHG simulations, additional RA-SHG datasets, and optical poling efficiency. "
}

\newpage
\textbf{Figure 1}\\

\newpage
\textbf{Figure 2}\\

\newpage
\textbf{Figure 3}\\

\newpage
\textbf{Figure 4}\\

\end{document}


\preprint{}

\title{Supplemental Material\\ \ \\
\normalsize Optical poling of a quantum ferroelectric metal across the order-disorder phase transition} 

\author{Mohamed Kandil}
\affiliation{Department of Physics, Auburn University, Auburn, AL 36849, USA}

\author{Yasuhide Tomioka}
\affiliation{National Institute of Advanced Industrial Science and Technology (AIST), Tsukuba 305-8565, Japan}

\author{Yafei Ren}
\affiliation{Department of Physics, University of Delaware, Newark, DE 19716, USA}

\author{Ryan Comes}
\affiliation{Department of Materials Science and Engineering, University of Delaware, Newark, DE 19716, USA}

\author{Isao H. Inoue}
\affiliation{National Institute of Advanced Industrial Science and Technology (AIST), Tsukuba 305-8565, Japan}

\author{Wencan Jin}
\email{wjin@auburn.edu}
\affiliation{Department of Physics, Auburn University, Auburn, AL 36849, USA}

\date{\today}

\maketitle

\tableofcontents

\newpage
\section{Synthesis and experimental methods}

\subsection{Sample synthesis}

\noindent \ce{Sr_{0.95}Ba_{0.05}Ti_{1-x}Nb_{x}O3} single crystals were synthesized using the floating-zone method as described in Ref.~\cite{Tomioka2022}. The resistivity and the Hall voltage for 5 K $\leq$ T $\leq$ 300 K were measured in the Physical Property Measurement System (PPMS, Quantum Design Inc.). The resistivity below 1 K was measured using an AC resistance bridge (LR700, Linear Research Inc.) in a cryostat using a $^3$He/$^4$He dilution refrigerator ($\mu$ dilution, Taiyo-Toyo Sanso Inc.). The temperature-carrier concentration phase diagram shown in Fig. 1(a) is constructed using the Hall mobility and resistivity data as reproduced in Fig. S1. 

\begin{figure*}[ht!]
    \centering
    \includegraphics[width=0.98\textwidth]{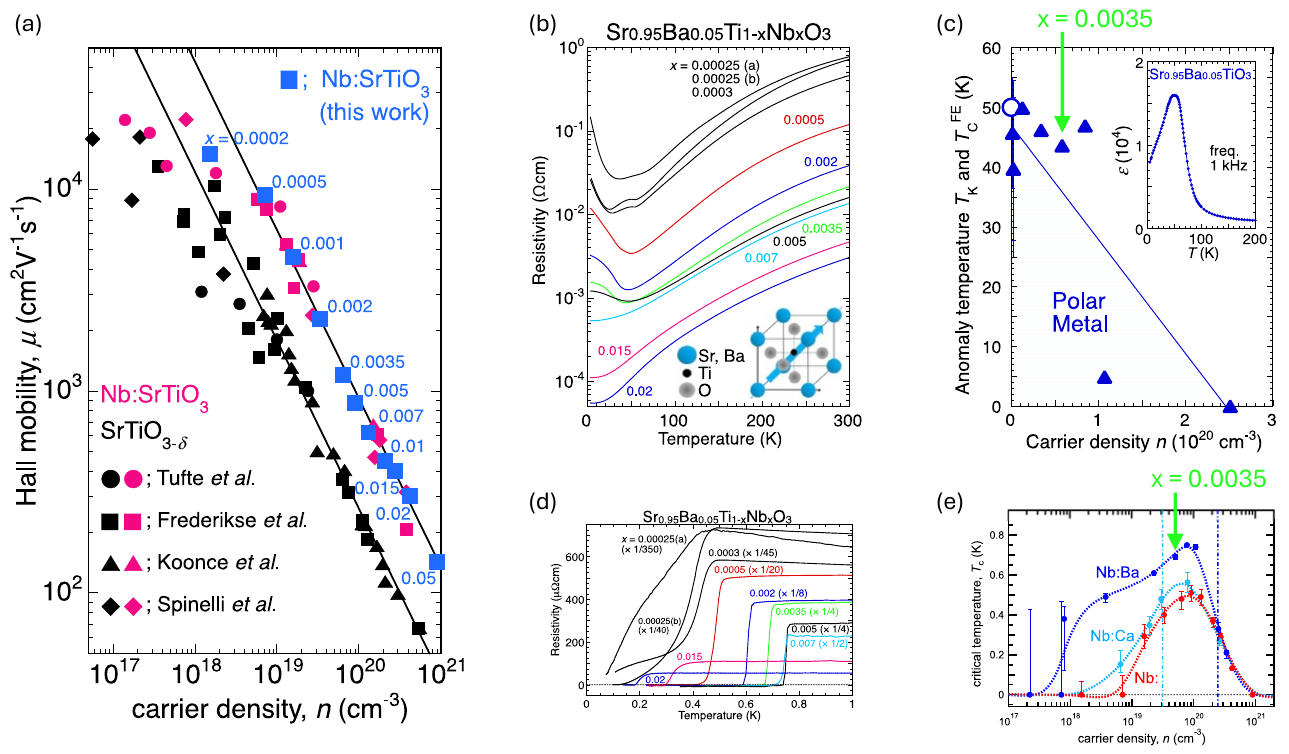}
    \caption{\label{fig:S1} \justifying \small \textbf{(a)} Hall mobilities of our Nb-doped STO samples at 5 K (blue symbols) in comparison with other reports in the literature. \textbf{(b)} Resistivity of our \ce{Sr_{0.95}Ba_{0.05}Ti_{1-x}Nb_{x}O3} single crystals with varying $x$. \textbf{(c)} The polar metal regime as defined by the resistivity anomaly temperature extracted from (b). \textbf{(d)} Resistivity of the \ce{Sr_{0.95}Ba_{0.05}Ti_{1-x}Nb_{x}O3} single crystals (the same samples that were studied in (b)) that show superconducting transitions below 1 K for varying $x$. \textbf{(e)} The superconducting dome as defined by the critical temperature extracted from (d). All the data were reproduced from Ref.~\cite{Tomioka2022}}
\end{figure*}

\subsection{Optical second harmonic generation} 

\noindent  A schematic of the RA-SHG experimental configuration can be found at Ref.~\cite{jin2020observation}. All the RA-SHG data were acquired using a normal incidence geometry, and the reflected SHG intensity was recorded as a function of the azimuthal angle between the incident electric polarization and the in-plane crystalline axis \textbf{a}. The incident ultrafast laser has 800 nm wavelength, $50\text{\,fs}$ pulse width, and $200\text{\,kHz}$ repetition rate, and was focused into a $50\,\mu  \text{m}$ spot in diameter using an achromatic lens. The SHG signal was detected using an EMCCD camera. In the pump-probe SHG experiments, a pump beam was added in the collinear geometry with the probe beam. The pump light source has $700\text{\,nm}$ wavelength, 70 fs pulse width, and $200\text{\,kHz}$ repetition rate, and was focused into a $80\,\mu \text{m}$ spot in diameter. Prior to the RA-SHG measurements, a fresh and flat surface was cleaved from the bulk crystal. The crystalline axes of the cubic phase were determined using x-ray diffraction at room temperature. The cleaved sample was mounted in a closed-cycle liquid helium cryostat, and all thermal cycles were carried out at a base pressure of 3$\times 10^{-7}$ Torr.

\newpage
\section{Characterization of electron-doped \ce{SrTiO3} at varying carrier concentrations}

\begin{figure}[ht!]
\includegraphics[width=0.98\textwidth]{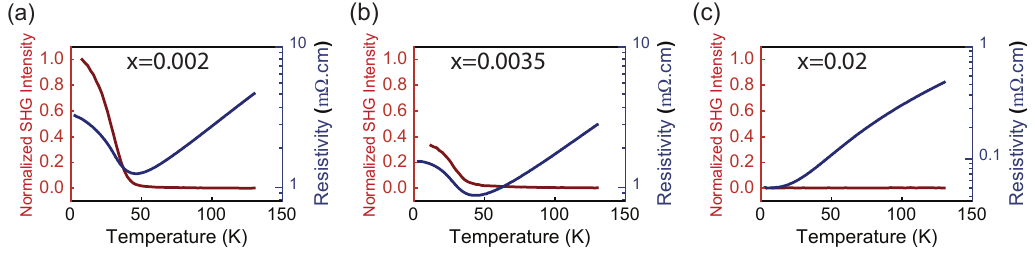}
\caption{\label{Fig:S2} \justifying \small Temperature dependence of SHG intensity (left axis) and electrical resistivity (right axis) for \ce{Sr_{0.95}Ba_{0.05}Ti_{1-x}Nb_{x}O3} bulk crystals with nominal Nb concentrations of (a) $x$ = 0.002, (b) $x$ = 0.0035, and (c) $x$ = 0.02.}
\end{figure}

\noindent Figure ~\ref{Fig:S2} shows the correlation between the onset of ferroelectricity and electric resistivity anomalies at varying carrier concentrations. The nominal \ce{Nb} concentrations of $x$ = 0.002, 0.0035, and 0.02 correspond to the Hall carrier concentrations of $2.27\times 10^{19}$ cm$^{-3}$, $5.04\times 10^{19}$ cm$^{-3}$, and $3.40\times 10^{20}$ cm$^{-3}$, respectively. From low to high doping concentrations, the dipolar interactions are gradually screened by charge carriers until long-range ferroelectric order is completely suppressed. 

\newpage
\section{Simulation of SHG in normal incidence geometry}

\subsection{4/mmm ($D_\mathrm{4h}$) point group}

\noindent \ce{SrTiO3} crystallizes in a cubic structure at room temperature and undergoes a cubic-to-tetragonal phase transition. In the low temperature phase of centrosymmetric 4/mmm ($D_\mathrm{4h}$) point group, the bulk electric quadrupole (EQ) SHG can be written as 

\begin{equation}
\label{eq:S1}
    P_{i}^{EQ}(2\omega)=\sum_{jkl}\chi_{ijkl}^{(3)}(D_\mathrm{4h})E_j(\omega)\nabla_{k} E_l(\omega)\\
\end{equation}

\noindent The $D_\mathrm{4h}$ point group allows 21 nonzero elements of $\chi_{ijkl}^{(3)}$ of which only 11 are independent:

\begin{center}
       xxxx = yyyy, zzzz,\\
    yyzz = xxzz,  yzzy = xzzx,  xxyy = yyxx, \\
    zzyy = zzxx,  yzyz = xzxz,  xyxy = yxyx, \\
    zyyz = zxxz,  zyzy = zxzx,  xyyx = yxxy. \\
\end{center}

\noindent One can write the susceptibility tensor as
\begin{equation}
\label{eq:S2}
\mathbf{\chi}_{ijkl}^{(3)} (D_\mathrm{4h}) =
\begin{pmatrix}
\begin{bmatrix}
\chi_{xxxx} & 0 & 0 \\0 & \chi_{yyxx} & 0 \\0 & 0 & \chi_{xxzz}
\end{bmatrix} &
\begin{bmatrix}
0 & \chi_{yxyx} & 0 \\ \chi_{yxxy} & 0 & 0 \\ 0 & 0 & 0
\end{bmatrix} &
\begin{bmatrix}
0 & 0 & \chi_{xzxz} \\ 0 & 0 & 0 \\ \chi_{xzzx} & 0 & 0
\end{bmatrix}
\\[3.5 em]
\begin{bmatrix}
0 & \chi_{yxxy} & 0 \\ \chi_{yxyx} & 0 & 0 \\ 0 & 0 & 0
\end{bmatrix} &
\begin{bmatrix}
\chi_{yyxx} & 0 & 0 \\ 0 & \chi_{xxxx} & 0 \\ 0 & 0 & \chi_{xxzz}
\end{bmatrix} &
\begin{bmatrix}
0 & 0 & 0 \\ 0 & 0 & \chi_{xzxz} \\ 0 & \chi_{xzzx} & 0
\end{bmatrix}
\\[3.5em]
\begin{bmatrix}
0 & 0 & \chi_{zxxz} \\ 0 & 0 & 0 \\ \chi_{zxzx} & 0 & 0
\end{bmatrix} &
\begin{bmatrix}
0 & 0 & 0 \\ 0 & 0 & \chi_{zxxz} \\ 0 & \chi_{zxzx} & 0
\end{bmatrix} &
\begin{bmatrix}
\chi_{zzxx} & 0 & 0 \\ 0 & \chi_{zzxx} & 0 \\ 0 & 0 & \chi_{zzzz}
\end{bmatrix}
\end{pmatrix}.
\end{equation} 

\noindent We simulate the RA-SHG patterns for all four polarization configurations (P$_\mathrm{in}$-P$_\mathrm{out}$, P$_\mathrm{in}$-S$_\mathrm{out}$, S$_\mathrm{in}$-P$_\mathrm{out}$, and S$_\mathrm{in}$-S$_\mathrm{out}$) in the functional forms:

\begin{equation}
\label{eq:S3}
\begin{aligned}
I_{PP}^{2\omega}(\varphi) &= \frac{\sin^{2}\theta}{16}
\Bigg[\Bigg(\cos^{3}\theta\Big(3\chi_{xxxx}-8\chi_{xzzx}+2\chi_{yxxy}+\chi_{yxyx}
+(\chi_{xxxx}-2\chi_{yxxy}-\chi_{yxyx})\cos4\varphi\Big) \\
&\quad +4\chi_{xzxz}\cos\theta\,\sin^{2}\theta
\Bigg)^{2} + \Bigg( \chi_{zxzx}\cos^{3}\theta +(-2\chi_{zxxz}+\chi_{zzzz})
\cos\theta\,\sin^{2}\theta \Bigg)^{2} \Bigg] \\
\\
I_{PS}^{2\omega}(\varphi) &= \frac{\sin^2{\theta}\cos^4{\theta}}{16}
\left( -\chi_{xxxx} + 2\chi_{yxxy} + \chi_{yxyx} \right)^2
\sin^2{4\varphi} \\
\\
I_{SP}^{2\omega}(\varphi) &= \frac{\sin^2{\theta}\cos^2{\theta}}{16}
\Bigg[ 16\chi_{xxxx}^2 + \Bigg (\chi_{xxxx} - 2\chi_{yxxy} + 3\chi_{yxyx} +\\
&\hspace{7cm}  (-\chi_{xxxx} + 2\chi_{yxxy} + \chi_{yxyx})\cos^2{4\varphi} \Bigg)^2 \Bigg] \\
I_{SS}^{2\omega}(\varphi) &= \frac{\sin^2{\theta}}{16}
\left( -\chi_{xxxx} + 2\chi_{yxxy} + \chi_{yxyx} \right)^2
\sin^2{4\varphi}
\end{aligned}
\end{equation}

\noindent For normal incidence geometry where $\theta = 0^{\circ}$, SHG intensity in all four polarization configurations vanishes.\\

\subsection{4mm (C$_\mathrm{4v}$) point group}

\noindent For strained \ce{SrTiO3} in noncentrosymmetric 4mm (C$_\mathrm{4v}$) point group, the bulk electric dipole (ED) SHG can be written as 

\begin{equation}
\label{eq:S4}
    P_{i}^{ED}(2\omega)=\sum_{jk}\chi_{ijk}^{(2)}(C_\mathrm{4v})E_j(\omega)E_k(\omega)\\
\end{equation}

\noindent The $C_\mathrm{4v}$ point group allows 7 nonzero elements of $\chi_{ijk}^{(2)}$ of which only 4 are independent:

\begin{center}
   xzx = yzy, xxz = yyz, zxx = zyy, zzz\\
\end{center}

\noindent One can write the susceptibility tensor as

\begin{equation}
\label{eq:S5}
\mathbf{\chi}_{ijk}^{(2)}(C_\mathrm{4v}) =
\begin{pmatrix}
\begin{bmatrix} 0 \\ 0 \\ \chi_{xxz} \end{bmatrix} &
\begin{bmatrix} 0 \\ 0 \\ 0 \end{bmatrix} &
\begin{bmatrix} \chi_{xxz} \\ 0 \\ 0 \end{bmatrix} \\[3.5em]

\begin{bmatrix} 0 \\ 0 \\ 0 \end{bmatrix} &
\begin{bmatrix} 0 \\ 0 \\ \chi_{xxz} \end{bmatrix} &
\begin{bmatrix} 0 \\ \chi_{xxz} \\ 0 \end{bmatrix} \\[3.5em]

\begin{bmatrix} \chi_{zxx} \\ 0 \\ 0 \end{bmatrix} &
\begin{bmatrix} 0 \\ \chi_{zxx} \\ 0 \end{bmatrix} &
\begin{bmatrix} 0 \\ 0 \\ \chi_{zzz} \end{bmatrix}
\end{pmatrix}
\end{equation}

\noindent We simulate the RA-SHG patterns for all four polarization configurations in the functional forms:

\begin{equation}
\label{eq:S6}
\begin{aligned}
I_{PP}^{2\omega}(\varphi) &= \sin^2{\theta}
\left[ 4\chi_{xxz}^2 \cos^4{\theta}
+ \left( \chi_{zxx}\cos^2{\theta} + \chi_{zzz}\sin^2{\theta} \right)^2 \right] \\
\\
I_{PS}^{2\omega}(\varphi) &= 0 \\
\\
I_{SP}^{2\omega}(\varphi) &= \chi_{zxx}^2 \sin^2{\theta}\\
\\
I_{SS}^{2\omega}(\varphi) &= 0 \\
\\
\end{aligned}
\end{equation}

\noindent For normal incidence geometry where $\theta = 0^{\circ}$, SHG intensity in all four polarization configurations vanishes.\\


\newpage
\section{Electric dipole contribution to SHG from polar nanoregions}

\noindent We derive the electric dipole contribution from the local dipoles \textbf{P} associated with the polar nanoregions: 

\vspace{-10pt}
\begin{equation}
\label{eq:S7}
    P_{i}^{ED}(2\omega)=\sum_{jk}\chi_{ijk}^{(2)}(\textbf{P})E_j(\omega)E_k(\omega)\\
\end{equation}

\noindent where the local polar vector $\mathbf{P} = ( \sin\alpha \cos\beta,\sin\alpha \sin\beta, \cos\alpha)$, and $\alpha$ and $\beta$ are angles with respect to the $z$-axis and $x$-axis, respectively. \\

\noindent Considering the local dipoles as a perturbation to the centrosymmetric tetragonal lattice, we use a Taylor expansion to express $\chi_{ijk}^{(2)}(\textbf{P})$ in linear terms

\vspace{-10pt}
\begin{equation}
\label{eq:S8}
\mathbf{\chi}_{ijk}^{(2)}(\mathbf{P}) =\sum\limits_{l} \frac{\partial \chi_{ijk}}{\partial P_l} P_{l} \text{ ,}
\end{equation}

\noindent One can define the third-order nonlinear susceptibility tensor $\chi_{ijkl}^{(3)}=\frac{\partial\chi_{ijk}}{\partial P_l}$, as expressed in the susceptibility tensor of $D_\mathrm{4h}$ point group in Equation (\ref{eq:S2}).\\

\noindent Combinging Equations (\ref{eq:S2}) and (\ref{eq:S8}), we obtain $\chi_{ijk}^{(2)}(\mathbf{P})$ 

\begin{equation}
\label{eq:S9}
\mathbf{\chi}_{ijk}^{(2)}(\mathbf{P}) =
\begin{pmatrix}
\begin{bmatrix} \chi_{xxxx} \cos \beta \, \sin \alpha \\ \chi_{yyxx} \sin \alpha \, \sin \beta \\ \chi_{xxzz} \cos \alpha \end{bmatrix} &
\begin{bmatrix} \chi_{yxyx} \sin \alpha \, \sin \beta \\ \chi_{yxxy} \cos \beta \, \sin \alpha \\ 0 \end{bmatrix} &
\begin{bmatrix} \chi_{xzxz} \cos \alpha \\ 0 \\ \chi_{xzzx} \cos \beta \, \sin \alpha \end{bmatrix} \\[3.5em]

\begin{bmatrix} \chi_{yxxy} \sin \alpha \, \sin \beta \\ \chi_{yxyx} \cos \beta \, \sin \alpha \\ 0 \end{bmatrix} &
\begin{bmatrix} \chi_{yyxx} \cos \beta \, \sin \alpha \\ \chi_{xxxx} \sin \alpha \, \sin \beta \\ \chi_{xxzz} \cos \alpha \end{bmatrix} &
\begin{bmatrix} 0 \\ \chi_{xzxz} \cos \alpha \\ \chi_{xzzx} \sin \alpha \, \sin \beta \end{bmatrix} \\[3.5em]

\begin{bmatrix} \chi_{zxxz} \cos \alpha \\ 0 \\ \chi_{zxzx} \cos \beta \, \sin \alpha \end{bmatrix} &
\begin{bmatrix} 0 \\ \chi_{zxxz} \cos \alpha \\ \chi_{zxzx} \sin \alpha \, \sin \beta \end{bmatrix} &
\begin{bmatrix} \chi_{zzxx} \cos \beta \, \sin \alpha \\ \chi_{zzxx} \sin \alpha \, \sin \beta \\ \chi_{zzzz} \cos \alpha \end{bmatrix}
\end{pmatrix}
\end{equation}

\vspace{15 pt}
\noindent We simulate the functional forms for the RA-SHG using normal incidence geometry in the parallel and crossed polarization configurations:

\vspace{-15 pt}
\begin{multline}
    I_{\parallel}^{2\omega}(\varphi) = \sin^2(\alpha) 
\Big[ (\chi_{xxxx} - \chi_{yxxy} - \chi_{yyxx} - \chi_{yxyx}) \, \sin(\beta - 3\varphi) + \\(3\chi_{xxxx} + \chi_{yxxy} + \chi_{yyxx} + \chi_{yxyx}) \, \sin(\beta + \varphi) \Big]
\end{multline}

\vspace{-15 pt}
\begin{multline}
I_{\perp}^{2\omega}(\varphi) = \sin^2(\alpha) 
\Big[ (-\chi_{xxxx} + \chi_{yxxy} + \chi_{yyxx} + \chi_{yxyx}) \, \sin(\beta - 3\varphi) + \\(\chi_{xxxx} + 3\chi_{yxxy} - \chi_{yyxx} - \chi_{yxyx}) \, \sin(\beta + \varphi) \Big]
\end{multline}

\noindent The RA-SHG patterns of $I_{\parallel}^{2\omega}(\varphi)$ and $I_{\perp}^{2\omega}(\varphi)$ are in a similar form, in which the amplitude of the SHG intensity is proportional to $\sin^2(\alpha)$, and the angular anisotropy is determined by $\beta$. Therefore, all the data we show in this work were acquired in the parallel polarization configuration unless otherwise specified.

\newpage
\section{Meltdown of the ferroelectric-like domain state with $\beta = -10^{\circ}$}

\begin{figure*}[ht!]
\includegraphics[width=0.99\textwidth]{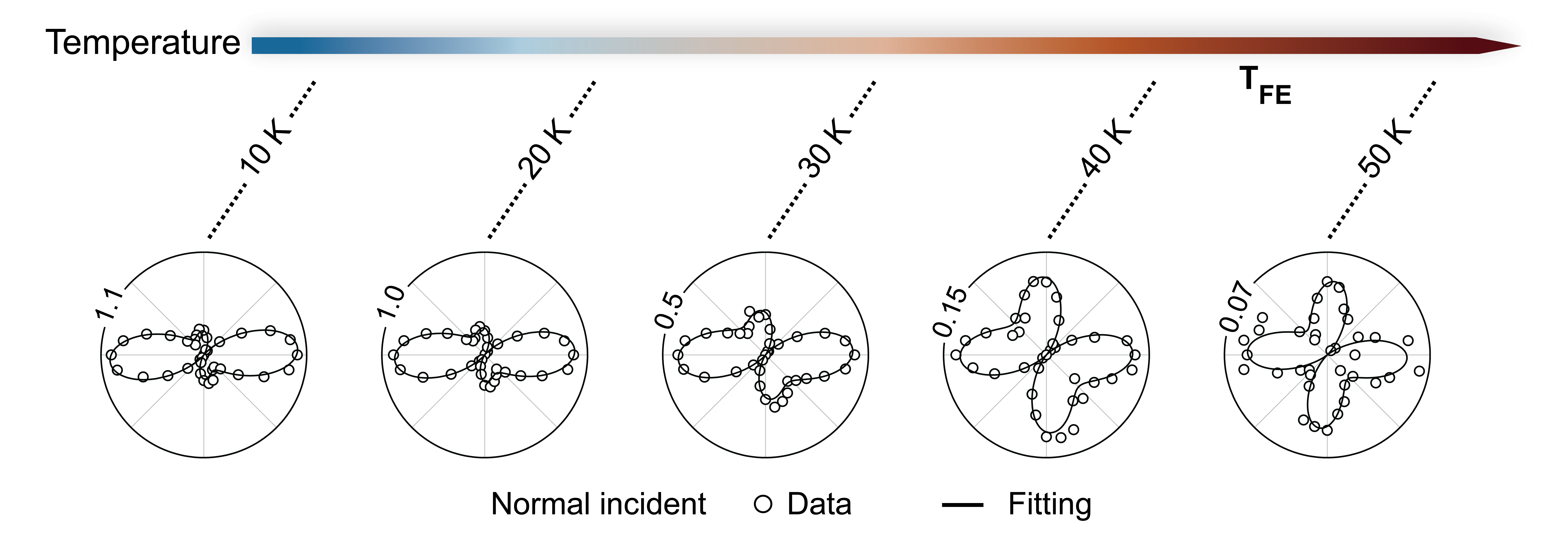}
\caption{\label{Fig:S3} \justifying \small Polar plots of RA-SHG data acquired in the
parallel channel of the normal incidence configuration at selected temperatures below and above $T_\mathrm{FE}$. All data sets are normalized to a value of 1 corresponding to the maximum SHG intensity at 10 K.}
\end{figure*}

\newpage
\section{Optical poling efficiency}

\begin{figure*}[ht!]
\includegraphics[width=0.6\textwidth]{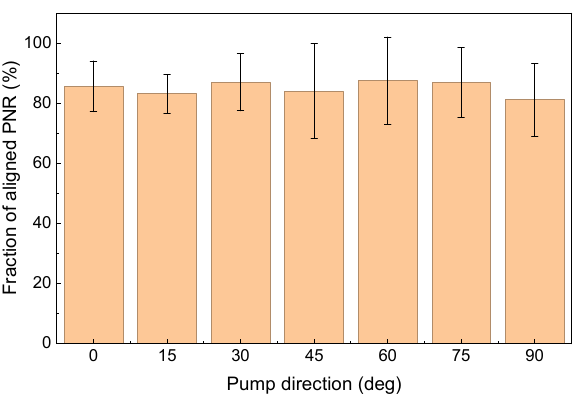}
\caption{\label{Fig:S4} \justifying \small Fraction of polar nanoregions (PNR) aligned with the pump optical field at varying pump directions.}
\end{figure*}

\noindent Above $T_\mathrm{FE}$, the polar nanoregions are randomly oriented, yielding the gray shaded RA-SHG patterns in Fig. 4 of the main text. The pump pulses align the polar nanoregions with the optical field, giving rise to pump-probe RA-SHG patterns. To determine the optical poling efficiency, we carry out simulations using 100 $\times$ 100 cells. We assume that by applying the pump pulses, $p\%$ of the polar nanoregions are aligned with the optical field, while the rest $1-p\%$ remain randomly oriented. We fit our simulation to the pump-probe RA-SHG patterns in Fig. 4 with pump directions along 0$^\circ$, 15$^\circ$, 30$^\circ$, 45$^\circ$, 60$^\circ$, 75$^\circ$, and 90$^\circ$. As shown in Fig.~\ref{Fig:S4}, over 80\% of the polar nanoregions are consistently aligned with the optical field, which represents a satisfactory optical poling efficiency. Even though our RA-SHG measurements do not directly resolve domains or domain walls, the fraction of aligned polar nanoregions extracted from our model indicates a strong pump-induced orientation bias of the polar nanoregions population. Domain formation and domain walls may affect the pinning and dynamics of this process, but addressing their roles requires spatially resolved experiments.

\newpage
\bibliographystyle{naturemag}
\bibliography{Supplementary.bib}
